\documentstyle[12pt,epsfig]{article}

 \newcommand {\be} {\begin{equation}}

\newcommand {\bea} {\begin{eqnarray} \nonumber }

\newcommand {\ee} {\end{equation}}

\newcommand {\eea} {\end{eqnarray}}

\newcommand {\cH}  {{\cal H}}

\newcommand {\by}  {{\bf y}}

\newcommand {\bx}  {{\bf x}}

\newcommand {\ap}  {\left(}

\newcommand {\cp}  {\right)}

\newcommand {\Tr} {\mbox{Tr}}

\newcommand {\sign} {\mbox{sign}}

\def \form#1 {eq. (\ref{#1}) }

\def \parziale#1#2  {{\partial {#1} \over \partial {#2}}}

\begin{document}

\title{A pedagogical introduction to the replica method for  glasses}

\author{ Giorgio Parisi \\
         Dipartimento di Fisica and INFN, Universit\`a di Roma {\sl La Sapienza},\\
        Piazzale Aldo Moro, Roma 00185, Italy }

\maketitle

\begin{abstract}
In this note I present a simplified version of the recent computation (M\'ezard and Parisi 1998, 
1999) of the properties of glasses in the low temperature phase in the framework of the replica 
theory, using an extension of the tools used in liquid theory.  I will only consider here the case 
of the internal energy at $T=0$, which can be studied in a simple way without introducing replicas.
\end{abstract}

\section{Introduction}

In this note I will present the recently developed theory, based on replicas, (M\'ezard and Parisi 
1998,1999) which aims to derive from first principles the basic thermodynamic properties of glasses 
using a modified liquid theory.  In order to simplify the presentation I am studying only one point: 
the computation of the ground state energy, i.e.  the internal energy at zero temperature ($E_{m}$).  
To this end I am going to use an explicit approach, in which the physical assumptions are evident.  
In this approach no replicas are introduced, but the same results can be obtained using the replica 
method.  This note can be considered a pedagogical introduction to the more sophisticated approach 
of M\'ezard and Parisi (1998, 1999).

Beyond technicalities the basic assumption is that {\sl a glass is very near to a frozen liquid}.  
This quite old statement can be rephrased as {\sl a liquid is very near to a heated glass}.  In 
other words the configurations of a glass at low temperature are not far from those of a liquid.  
Therefore if we use some smart method to explore the phase space in the liquid phase we can find the 
properties in the low temperature glassy phase.

This strategy has been put into action using the replica theory.  In this way one finds, as output 
of an explicit computation, that the glass transition is characterized by the vanishing of the 
configurational entropy (the so called complexity) and in the replica language the low temperature 
phase is described by one step replica symmetry breaking (M\'ezard et al.  1987, Parisi 1992, 
Kirkpatrick and Thirumalai 1995) with a non vanishing non-ergodic parameter at the phase transition 
point.  The computations can be done at all temperatures.  Here we are going to reproduce the 
results of the replica approach at $T=0$, where only energetic considerations (not entropic ones) 
are relevant.

\section{Basic assumptions}

The strategy of the computation is rather simple.  Given a system of $N$ particles with an 
Hamiltonian $H(\bx)$, we would like to know the minimum value of $H$.  This task in non trivial when 
$N \to \infty$ if the ground state is not translationally invariant.  The minimum can be 
characterized as the solution of the stationary equations

\be
{\partial H \over \partial x_{i}} \equiv F_{i} =0\label{STATIC},
\ee
having the minimum energy density $E \equiv H/N$.

In order to put the problem in a wider perspective it is convenient to 
consider the function

\be \nu(E)\equiv \sum_{\alpha}\delta (E-E_{\alpha}) =\int d\bx \ 
|\det(\cH (\bx)| \delta (NE -H(\bx)) \prod_{i=1}^{N}\delta (F_{i}(\bx)) ,
\ee 
where the sum runs over all the solutions of eq (\ref{STATIC}) and the Hessian matrix $\cH$ is an $N 
\times N$ matrix defined as

\be 
\cH_{i,k}={\partial H \over \partial x_{i} x_{k}}.
\ee

There are many arguments which imply that in the limit 
$N$ going to infinity we have (in distribution sense)

\be
\nu (E) \approx \exp (N \sigma(E)),
\ee
where the function $\sigma(E)$ is positive for $E>E_{m}$, and vanishes at $E=E_{m}$: $\nu(E)=0$ for 
$E<E_{m}$.  Therefore our goal is reached if we evaluate the function  $\sigma(E)$.

Unfortunately  the computation  of $\sigma(E)$ may be too complicated. In many cases 
it is convenient to define the function

\be
\mu(E)\equiv \sum_{a}\delta (E-E_{a})\ \sign(\det(\cH)(\bx)))
 =\int d\bx \
\det(\cH (\bx) \delta (NE -H(\bx_{a})) \prod_{i=1}^{N}\delta (F_{i}(\bx)).
 \ee

The advantage of this second definition is that the final integrand is a
smooth function of
$\bx$ and this may simplify the computation. 

In the same way we define in the large $N$ limit the function

\be
\mu(E) \approx \exp (N \Sigma(E)).
\ee
Of course in the region where the Hessian is positive the two
definitions coincide.

In the region of $E$ near to $E_{m}$ we expect that most of the stationary points are 
minima and therefore the function $\sigma(E)$ and $\Sigma(E)$ are similar in this region 
(Cavagna et al.  1998).  As a consequence both functions can be used for the computation 
of the ground state.  In the rest of the paper we are going to compute $\Sigma(E)$, 
however within the approximation we will do in this note the two functions coincide.

The evaluation of $\Sigma(E)$ may be simplified (Monasson 1995) if we 
introduce the generalized partition function

\be
\zeta(\gamma)
=\int dE \exp (-N (\gamma E -\Sigma(E))) =
\int d\bx \ \det(\cH(\bx)) \exp (- \gamma H(\bx)) \prod_{i=1}^{N}
\delta (F_{i}(\bx)).
\ee

In the large $N$ limit we have that

\bea
\zeta(\gamma)= \exp(-N \Phi(\gamma)),\\
\Phi(\gamma)= \gamma E(\gamma) -\Sigma(E(\gamma)).\\
E(\gamma) ={\partial \Phi(\gamma) \over \partial \gamma}. \nonumber
\eea

The previous equations are the equivalent of the usual relation among the free energy and the entropy.  
The quantity $\Sigma$ is the equivalent of the entropy and is called the complexity or the 
configurational entropy.

We now face the problem of computing the quantity $\Phi(\gamma)$.  We are clearly interested in the 
region of relatively large $\gamma$, which corresponds to small energies and consequently small 
$\Sigma$.  The computation of $\Phi(\gamma)$ is rather involved.  In order to simplify it we can 
make three assumptions on the configurations which dominate the integral at large $\gamma$, that are 
similar to the configurations which dominate the usual partition function at large $\beta$.

\begin{itemize}

\item The integrand is dominated by those configurations in which all the eigenvalues are positive, 
i.e by configurations near to a minimum, not to a saddle point.

\item The dominant integration region splits into many, approximately disjoint, regions (${\cal 
R}_{\alpha}$) and in each of these regions the Hamiltonian can be well approximated by a quadratic 
one.

\item The variations of the determinant of the Hessian from one region to an other region can be 
neglected.  Therefore the determina
nt of the Hessian is nearly constant in all the regions.

\end{itemize}

\section{How to use the assumptions}

We can now proceed. We use the following representation of
the $\delta$ function:

\be
\delta (F_{k})= \int d \lambda_{k} \exp (i\lambda_{k}F_{k}).
\ee
We thus find

\be
\zeta(\gamma)=
\int d\bx d{\bf \lambda}  \det(\cH(\bx)) \exp \ap - \gamma H(\bx)
 +i \sum_{k }\lambda_{k}F_{k}\cp.
\ee

If we now shift the integration over the $x$ variables and we introduce the variables
$y_{k}=x_{k}+i \gamma^{-1}\lambda_{k}$ we obtain

\be
\zeta(\gamma)=
 \int d\by d{\bf \lambda} \det(\cH(\by)) \exp \ap- \gamma H(\by)  -
  \sum_{k,i }\lambda_{k}\lambda_{i}\cH(\by)_{k,i} +O(H^{III})\cp.\label{ZETA}
\ee

We can now neglect the anharmonic terms proportional to the third derivative of $H$, (i.e.  the 
terms of order $H^{III}$) to perform the integral over the $\lambda$.  We arrive at the simpler 
expression

\be
\zeta(\gamma)=
  \int d\by  \exp (- \gamma H(\by) +\Phi^{H}(\gamma,\by)), \label{FINAL}
 \ee
where 

\bea
\exp(-\Phi^{H}(\gamma,\by)) =\int d{\bf \lambda} \exp \ap-
\gamma\sum_{k,i }\lambda_{k}\lambda_{i}\cH(\by)_{k,i}
\cp=\\
\ap {2 \pi \over \gamma}\cp^{3N/2} \det (\cH)^{-1/2}= 
\ap{2 \pi \over \gamma}\cp ^{3N/2}  \exp \ap-1/2\Tr( \ln (\cH))\cp.
\eea

The formula (\ref{FINAL}) can be derived directly from the previous assumptions.  Indeed if we start 
from eq.  (\ref{FINAL}) and neglect the anharmonic terms in each region, we find that the 
determinant is constant so it can be taken out of the integration and we obtain:

\be 
\sum_{\alpha}\exp(-\Phi^{H}_{\alpha}(\gamma))\int_{{\cal R}_{\alpha}}
d\by  \exp(-\gamma H(\by))=
\sum_{\alpha} \exp (-\gamma E)_{\alpha}.
\ee 

Indeed the integral in each of the regions ${\cal R}_{\alpha}$ in which the integrand is 
approximately harmonic gives approximately $\exp (-\gamma 
E_{\alpha}+\Phi^{H}_{\alpha}(\gamma))$ (Angelani et 
al.  1998).

Although the formula (\ref{FINAL}) can be derived directly, the previous detailed proof is useful in 
order to understand which are the possible steps that must be followed in order to improve its 
validity if some of the assumptions are removed.

Also after these simplifications the evaluation of eq.  (\ref{FINAL}) is not trivial because the 
determinant changes from point to point.  Assuming that it takes the same value near all the low 
lying minima, we get
\be \zeta(\gamma)=Z(\gamma)\exp\ap \Phi^{H}(\gamma) \cp=\exp\ap-\gamma 
F(\gamma)+ \Phi^{H}(\gamma)\cp,
\ee
where $\Phi^{H}(\gamma)$ is the average value of the harmonic free energy (multiplied by 
$\gamma$) for a configuration at 
temperature $1/\gamma$ and $Z(\gamma)$ is the usual partition function at $\beta=\gamma$.  We need 
to know only the free energy ($F(\gamma)$) and the spectral density of the INN (instantaneous normal 
modes, i.e.  the eigenvalues of the Hessian) in the liquid phase as a function of the temperature in 
order to find $\zeta(\gamma)$.

Let us use the previous formulae for $\Phi^{H}(\gamma)$ and let us introduce the corresponding 
harmonic entropy:
\be
S^{H}(\gamma)=-\Phi^{H(\gamma)}+\frac32\equiv-\frac32 \ln(\gamma)+\Delta(\gamma),
\ee
where $\Delta(\gamma)$ depends on $\gamma$ only via the eigenvalues of the Hessian in the 
liquid at a temperature $\gamma^{-1}$.
After a simple computation we find that the complexity is given by
\be
\Sigma(\gamma)=S^{L}(\gamma)-S^{H}(\gamma)+\gamma{d \Delta(\gamma)\over d\gamma},\label{EUREKA}
\ee
where $S^{L}(\gamma)$ is the entropy in the liquid.

In this way we get a simple expression for the complexity in terms of quantities that are 
defined in the liquid.  We need to known then in the high temperature phase down to a 
temperature $1/\gamma^{*}$, where $\Sigma(\gamma^{*})=0$.

The zero temperature configurations of the glass will not be the same as those of the liquid at 
finite temperature.  In the harmonic approximation we have to find the minima by assuming a 
quadratical energy and estimating the position of the particles in a glass using the formula
\be
y_{i}=x_{i}+\delta_{i},
\ee
where $x_{i}$ and $y_{i}$ are the position of the particles in the minimum and in the 
liquid respectively and 
\be
\sum_{k=1,N} \cH_{i,k}\delta_{k}=F_{i}.\label{NOBILITATE}
\ee

The analytic solution of the equation (\ref{NOBILITATE}) is not so simple, and it will not be 
discussed here.

\section{How to implement previous ideas}

It is always possible to evaluate eq.  (\ref{FINAL}) using numerical simulations in the 
liquid phase (Coluzzi et al 1999a and 1999b).  This is by far the simplest method.  I 
now present a possible analytic implementation of this strategy (M\'ezard and Parisi 1998, 
1999).

The computation of the free energy in the liquid phase does not present any serious difficulty.  
There are many methods which lead to integral equations for the correlation functions (e.g.  the 
hypernetted chain approximation, the mean spherical approximation, the Percus Yevick approximation).  
We have used the hypernetted chain approximation, however the other approximations may work better.  
In any case a clean computation of the properties of the liquid is standard (in the case of two body 
central interaction).

The computation of the spectrum is much more difficult, Most of the computations are done at low 
density and in spite of this simplification the results obtained are approximated (for example the 
localized tail at small and large eigenvalues is lost).  The only exact computation at low density, 
which corresponds to the first order in the virial expansion, has just been done recently (Cavagna 
et al 1999).  Lacking for the moment a systematic way to extend this computation to the high density 
region, we are following a quite simple method which becomes exact in some limit.  A closed formula 
can be obtained if we assume that the range of the Hessian is much larger than the typical 
interatomic distance and we use the superposition approximation
\be
g(x_{1}x_{2}x_{3})=g(x_{1}-x_{2})g(x_{2}-x_{3})g(x_{3}-x_{1})
\ee
in the cases where it is needed (M\'ezard and Parisi 1998, M\'ezard et al.  1999).

In order to compute the eigenvalues of the Hessian we compute analytically the moments of the 
eigenvalue distribution

\be M_{k}=\Tr (\cH ^{k}).
\ee 

Each moment can be written as the appropriate integral over the correlations functions.  In the long 
range approximation some terms are dominant over the others.  We keep only those terms and we 
introduce the superposition approximation into the appropriate places.  We use the large range 
expansion to select the diagrams.  The result we obtain for the spectrum is just a first 
approximation, the long range expansion is a nice way to organize the diagrams and we plan to 
improve it in the next future.  An unified approach, which interpolates in a reasonable way the 
expressions in the low density with those in the high density limit would be very useful.

After some computations (which we do not report here) we find a spectral density $\rho(e)$ that has 
support only in the region of positive eigenvalues $e$ (real frequencies $\omega=\sqrt(e)$) and goes 
to zero as $\omega$ as expected.  Some extra work is needed to improve the approach and having a 
realistic $S(k,\omega)$ at $T=0$.

We have applied this formalism to soft spheres with repulsive interaction (M\'e\-zard and 
Parisi 1998, 1999), to binary mixtures of soft spheres (Coluzzi et al.  1999a) and to 
binary mixture of particle with LJ potential (Coluzzi et al.  1999b).  We plan to study in 
the future undercooled Argon and $Si_{2}O_{4}$.

\begin{figure}
\centerline{\hbox{
\epsfig{figure=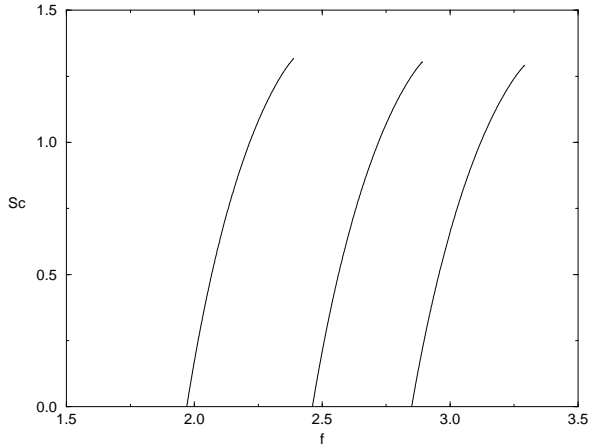,width=7cm,angle=-90}
}}

\caption{The configurational entropy $\Sigma(f)$ versus the free energy, (from M\'ezard and Parisi 
1999) computed within the harmonic resummation, at temperatures $T=0,.05,.1$ (from left to right).  
} \label{I} \end{figure}

Here I present the results at $T=0$ for  soft spheres with a repulsive
$x^{-12}$ potential.  In fig.  \ref{I} we see the complexity as function of free energy at three 
different temperatures (at zero temperature the free energy become identical to the internal 
energy).  The function $\Sigma$ should be considered reliable at low energy, however at high energy 
the harmonic approximation fails and non linear terms become more important.  Moreover in the real 
world increasing the energy the number of negative (maybe localized) eigenvalues increases and our 
approximation becomes no more reliable.

If we extend the same arguments from energy to free energy (following Monasson 1995) we must 
introduce replicas.  We find that there is transition at temperature $T_{K}$ such that:

\begin{itemize}

\item

For temperature greater than $T_{K}$ the number of minima which can be reached by slowly cooling the 
equilibrium configurations increase as $\exp(N\Sigma(E,T))$.  The equilibrium configurational 
entropy goes to zero at $T_{K}$.
\item 

For temperature smaller that $T_{k}$ there is a finite number of of minima which can be reached by 
slowing cooling the equilibrium configurations.  The configurational entropy is zero below $T_{K}$.

\item
The transition temperature corresponds to a value $\gamma_{K}$ which satisfies the 
following relation (M\'ezard and Parisi 1998)
\be
\Sigma^{L}(\gamma)=\Sigma^{H}(\gamma)
\ee
We notice that, if we neglect the temperature dependence of $\Delta(\gamma)$, this 
equation coincides with the condition $\Sigma(\gamma)=0$. The value of
$\gamma_{K}$ coincides with the value of $\gamma^{*}$ which is used to compute the ground 
state energy.

\end{itemize}

This is an explicit realization of a transition driven by an entropy crisis, which is the first 
simple realization in the random energy model (Derrida 1980).  The specific heat has a jump 
downward, when we decrease the temperature, which is the opposite of the typical behaviour in 
transitions characterized my the onset of a conventional order parameter (ferromagnets, 
superconductors, have a jump upward when we decrease the temperature).

\section{Conclusions}

We have found a  method which is able to use liquid theory method in 
the glasses phase putting in practice the old  adagio {\sl a glass is  a frozen 
liquid}.
 
We are able to compute with a reasonable approximation the thermodynamics and with a little more 
effort we can compute the static and the dynamic structure functions.

I would finally stress that in this theoretical framework we do not only obtain reasonable results 
for conventional experiments, but we have also clear and quantitative predictions for the relations 
between the response and the correlations in the aging regime (Cugliandolo and Kurchan 1993, Franz 
and M\'ezard 1994, Franz et al.  1998, 1999).  An experimental test of these relations, which can be 
done with present technology but with a lot of ingenuity, would have extremely deep consequences for 
theory (this theory would be just killed in the case of a negative result, while the whole approach 
based on replicas would be confirmed in the positive case).  These violations have been seen quite 
clearly in numerical simulations (Parisi 1997a and 1997b) and the time is ripe for seeing them in 
real experiments.

I hope that the experimentalists will consider such a measurement as a challenge and the results 
will be available for the next Andalo meeting.

\section*{Acknowledgments} It is a great pleasure for me to thank Marc M\'ezard for the long 
collaboration on the study of glassy systems which is at the root of the results presented in this 
paper.  I would also thank A.  Cavagna and I.  Giardina for very useful discussions.

\section*{References}

$\ \ \ \ $ L.  Angelani, G.  Parisi, G Ruocco and G Viliani (1998) Phys.  Rev.  Lett.  {\bf 81} 
4648.

A. Cavagna, I. Giardina and G. Parisi 1998, Phys. Rev. {\bf B57} 11251.

A.  Cavagna, I.  Giardina and G.  Parisi 1999, {\it Analytic computation of the Instantaneous Normal 
Modes spectrum in low density liquids} cond-mat preprint 9903155, Phys. Rev. Lett. to be published.

B.  Coluzzi, M.  M\'ezard, G.  Parisi and P.  Verrocchio 1999a, {\it Thermodynamics of binary 
mixture glasses}, cond-mat preprint 9903129, submitted to J.  Chem.  Phys..

B.  Coluzzi, G.  Parisi and P.  Verrocchio 1999b, {\it Lennard-Jones binary mixture: a 
thermodynamical approach to glass transition}, cond-mat preprint 9904124, submitted to J.  Chem.  
Phys..

L. Cugliandolo and J. Kurchan 1993, Phys.  Rev.  Lett.  {\bf 71}, 1.

B. Derrida 1981, Phys. Rev. {\bf B24}  2613.

S. Franz and M. M\'ezard 1994, Europhys.  Lett.  {\bf 26}, 209.

S.  Franz, M.  M\'ezard, G.  Parisi and L.  Peliti 1998, Phys.  Rev.  Lett.  {\bf 81} 1758.

S.  Franz, M.  M\'ezard, G.  Parisi and L.  Peliti 1999, {\it The response of glassy systems to 
random perturbations: A bridge between equilibrium and off-equilibrium} cond-mat preprint 9903370, 
submitted to J.  Stat.  Phys..

T.R.  Kirkpatrick and D.  Thirumalai 1995, Transp.  Theor.  Stat.  Phys.  {\bf 24} 927 and 
references therein.

M.  M\'ezard, G.  Parisi and M.A.  Virasoro 1987, {\sl Spin glass theory and beyond}, World 
Scientific (Singapore).

M. M\'ezard and G. Parisi 1998, Phys.  Rev. Lett. {\bf 82} 747.

M.  M\'ezard and G.  Parisi 1999, {\it A first principle computation of the thermodynamics of 
glasses} cond-mat preprint 9812180, J.  Chem.  Phys.  in press.

R. Monasson 1995, Phys. Rev. Lett. {\bf 75}  2847.

G.  Parisi 1992, {\sl Field Theory, Disorder and Simulations},
World Scientific, (Singapore).

G. Parisi 1997a, Phys. Rev. Lett., {\bf 79} 3660.

G. Parisi 1997b, Phil. Mag. {\bf B77} 257.

\end{document}